**Rewritable Complementary Nanoelectronics Enabled by Electron-Beam Programmable Ambipolar Doping**

Qing Lan[1,2#], Wenqing Song[1,2#], Siyin Zhu[1,2], Yi Zhou[1,2], Lu Wang[1,2], Junjie Wei[3], Jiaqi Liu[1,2], Zejing Guo[1,2], Takashi Taniguchi[4], Kenji Watanabe[5], Hai Huang[6], Jingli Wang[3], Xiaodong Zhou[1,7*], Alex Zettl[8], Jian Shen[1,2,7,9], Wu Shi[1,7*]

[1]*State Key Laboratory of Surface Physics and Institute for Nanoelectronic Devices and Quantum Computing, Fudan University, Shanghai 200433, China*

[2]*Department of Physics, Fudan University, Shanghai 200433, China*

[3]*College of Integrated Circuits & Micro-Nano Electronics, Fudan University, Shanghai 200433, China*

[4]*Research Center for Materials Nanoarchitectonics, National Institute for Materials Science, 1-1 Namiki, Tsukuba 305-0044, Japan*

[5]*Research Center for Electronic and Optical Materials, National Institute for Materials Science, 1-1 Namiki, Tsukuba 305-0044, Japan*

[6]*Shanghai Frontiers Science Research Base of Intelligent Optoelectronic and Perception, Institute of Optoelectronic and Department of Material Science, Fudan University, Shanghai 200433, China*

[7]*Zhangjiang Fudan International Innovation Center, Fudan University, Shanghai 201210, China*

[8]*Department of Physics, University of California, Berkeley, California 94720, USA*

[9]*Hefei National Laboratory, Hefei 230088, China*

[#] These authors contributed equally to this work.

[*] Corresponding author. E-mail: zhouxd@fudan.edu.cn (X.D.Z.); shiwu@fudan.edu.cn (W. S.).

**The ability to reversibly and site-selectively tune ambipolar doping in a single semiconductor is crucial for reconfigurable electronics beyond silicon, but remains highly challenging. Here, we present a rewritable architecture based on electron-beam programmable field-effect transistors (FETs). Using $WSe_2$ as a model system, we demonstrate electron-beam-induced doping that enables reversible, precisely controlled carrier modulation exceeding $10^{13}$ $cm^{-2}$. The *in-situ* writing, erasing, and rewriting of ambipolar doping of nanoscale patterns was directly visualized by scanning microwave impedance microscopy. This mask-free, lithography-compatible approach can achieve precise band engineering within individual channels, yielding near-ideal subthreshold swings (~ 60 mV/dec) and finely tunable threshold voltages for both carrier types without specialized contact engineering. These capabilities allow on-demand realization of high-performance logic, including CMOS inverters with high voltage gains and low power consumption, as well as NAND-to-NOR transitions on the same device via direct polarity rewriting. Our platform offers a scalable and versatile route for rapid prototyping of complementary electronics.**

Patterning doping profiles with nanometer precision is the cornerstone of modern semiconductor electronics. In silicon, this capability is achieved through ion implantation, which robustly defines complementary regions and enables scalable CMOS technology. Recently, two-dimensional transition metal dichalcogenides (TMDs) combine atomic-scale thickness, sizable bandgaps, and excellent electrostatic gate control, have emerged as promising candidates to succeed silicon in deeply scaled CMOS technology[1,2]. From basic inverters[3–7] to a 32-bit RISC-V microprocessor built from 5,900 $MoS_2$ transistors[8], recent advances have demonstrated the feasibility of wafer-scale growth and device integration[5,9–12].

However, extending site-selective, reversible, and nanoscale doping control to emerging 2D materials remains a major challenge. Existing approaches to TMD-based logic circuits offer only partial solutions, but fall short of simultaneously achieving reversibility, high spatial precision, scalability, and integration simplicity–requirements that are essential for realizing a truly complementary architecture. They typically rely on unipolar *n*-MOS logic[3,4,6–8,11] or the integration of distinct intrinsically *p*- and *n*-type TMDs[10,13,14], approaches that compromise the energy efficiency, layout density, and design simplicity afforded by a monolithic 2D CMOS platform. Unlocking the full potential of 2D electronics thus requires the ability to locally pattern both *p*- and *n*-type regions within a single TMD, analogous to ion implantation in silicon, with nanometer-scale precision and without comprising the channel.

A variety of doping strategies has contributed to progress in this direction[15–35], yet none fully satisfies the stringent requirements for scalable and reconfigurable 2D complementary electronics. Substitutional or ion implantation doping[15–21] provides permanent carrier modulation but often introduces lattice damage[2,36], strain[15], and fabrication complexity[18,21]. Charge-transfer doping via surface adsorbates[22–25], metals/metal oxides[26–31], or molecular overlayers[22,30,32–35], offers a non-destructive alternative but is generally limited by static doping profiles and micrometer-scale spatial resolution[26,27,29–31]. Electrostatic approaches employing split gates[37–40] or ferroelectric layers[37,41], enable dynamic polarity switching[37–40] and higher resolution[41], but add structural complexity and integration challenges[25]. Moreover, most existing schemes rely on dual-metal contacts[9,20,29–31,42–44] or multiple TMDs[10,13,14,44] to achieve both polarities, thereby limiting reconfigurability and increasing process cost. Realizing well-balanced *n*- and *p*-channel performance with precise threshold voltage ($V_{TH}$) control remains a critical prerequisite for optimized CMOS operation, which current doping strategies generally fail to achive[45,46]. Therefore, a truly rewritable method that enables on-demand, damage-free patterning of individual TMD into *p*- or *n*-type regions with nanometer-scale precision is highly desirable for realizing fully complementary and reconfigurable 2D electronics.

Here, we demonstrate a rewritable, electron-beam programmable electronics platform built from ambipolar $WSe_2$ field-effect transistors (FETs), which can be locally configured in situ into *p*-FETs, *n*-FETs, rectifiers, inverters, and fully functional CMOS logic gates. By doping selective regions of the $WSe_2$ channel via controlled e-beam writing, and characterizing the resulting carrier profiles through scanning microwave impedance microscopy (sMIM), we achieve reversible *p*- and *n*-type doping with ~ 200 nm precision. This precise control allows dynamic band engineering within individual channels to optimize device metrics, enabling near-ideal subthreshold swing of ~ 60 mV/dec and tunable threshold voltages for both polarities without the need for complex contact engineering. These capabilities allow on-demand design and construction of high-performance logic units, such as CMOS inverters with tunable switching voltages, optimized voltage gains and low power consumption. We also demonstrate seamless logic transitions between NAND and NOR gates on the same device via e-beam-enabled polarity rewriting, without physical reconstruction. Coupled with nanoscale resolution, scalability and ubiquity of e-beam lithography, this rewritable architecture offers a scalable, sustainable route for rapid prototyping and development of high-performance reconfigurable 2D complementary electronics.

**Rewritable 2D electronics platform via controllable e-beam doping**

Figure 1a illustrates our e-beam-programmable electronics platform based on 2D FETs for reconfigurable nanocircuits. Using a scanning e-beam, we can selectively "write" nonvolatile *p*- or *n*-doped regions, analogous to conventional ion implantation. The high-resolution e-beam writes can produce nanoscale doping profiles, demonstrated by the sMIM images of e-beam written line, triangular and letter patterns with minimal line-width of ~200nm in Fig.1a (see Supplementary Note 1 and Fig. S1). Direct rewriting of both doping polarities on the same device enables in-situ fabrication and optimization of basic logic devices like FETs and diodes with tunable threshold voltages and switching behavior. These devices can be dynamically integrated into complex circuits for rectification, amplification, or various logic functions. Crucially, the same device can be "erased" and "rewritten" with a different doping profile, allowing complete logic reconfiguration without physical alteration. This rewritable capability enables in-situ circuit reprogramming, significantly accelerating prototyping while reducing material waste, particularly valuable for increasingly complex circuit architectures.

To realize this rewritable platform, we employ ambipolar $WSe_2$ channels encapsulated in hexagonal boron nitride (hBN), which stabilizes the e-beam-induced doping and protects the semiconductor during e-beam exposure[47–49]. The 3D illustration in Fig.1a shows the device structure, where the heterostructures are assembled via standard polycarbonate (PC)-assisted dry-try transfer onto dielectric substrates ($HfO_2$, $Al_2O_3$, or $SiO_2$). Metal contacts (3nm Cr/70nm Au) are patterned using standard e-beam lithography (EBL) without additional contact engineering, making the process broadly compatible. All e-beam doping is performed in situ inside a scanning electron microscope (SEM) equipped with electrical feedthroughs, enabling real-time transport monitoring. By tuning the gate-setting voltages ($V_G = V_{SET}$) during e-beam exposure, we precisely control local doping density and polarity, following the same procedures and similar mechanism as previously demonstrated in graphene and $MoS_2$ devices[50], as illustrated in Supplementary Fig. S2.

To evaluate the tunability and generality of e-beam-induced ambipolar doping, we fabricated hBN-encapsulated $WSe_2$ devices on $Al_2O_3$ and $HfO_2$ substrates (Supplementary Fig. S3 and S4). Figure 1b shows transfer curves for a typical device on 50 nm $Al_2O_3$ at a drain-source bias $V_{DS}$ = 0.1 V, measured after sequential e-beam "erasing" and "rewriting" under 1 keV e-beam exposure at different $V_{SET}$. In-situ electrical measurements confirm reversible polarity switching between *n*-type and *p*-type dominant FET behavior by simply adjusting $V_{SET}$ during e-beam exposure. Similar behavior was observed in devices on $HfO_2$ and $SiO_2$ (Supplementary Fig. S4 and S5), demonstrating the stable non-volatile doping effects and the universality of our approach across various dielectric conditions. Extracted threshold voltages $V_{TH}$ from transfer curves in Fig. 1b exhibit a linear dependence on $V_{SET}$ for both polarities (Fig. 1c), underscoring the precision and reproducibility of this e-beam doping approach. This result suggests that e-beam doping enables well-balanced *n*- and *p*-FET performance as well as multi-$V_{TH}$ optimization for CMOS applications. Furthermore, we estimated from the large shift of threshold voltages and calculated the carrier density change $\Delta n$ in the same device (Fig. 1d), demonstrating strong carrier tunability across a wide range over ± $10^{13}$ $cm^{-2}$ for both *n*- and *p*-doping. This combination of reversibility, precision, and wide-range carrier control for both polarities, typically unattainable with conventional methods, establishes $WSe_2$ FETs as a viable platform for realizing rewritable 2D complementary electronics.

**Gate-tunable *p-n* junction via spatially selective e-beam patterning**

The scanning e-beam in a standard SEM system can be precisely directed to target regions using lithography functions, enabling spatially resolved doping with sub-micron accuracy. To demonstrate this capability, we patterned a lateral *p–n* junction within a single $WSe_2$ device by sequential e-beam exposures and evaluated its performance as a gate-tunable rectifier. As illustrated in Fig. 2a, the channel was initially globally doped *n*-type by 1 keV e-beam exposure at $V_G = V_{SET1}$. Subsequently, half of the channel was selectively re-doped under a different back-gate bias at $V_{SET2}$, converting that region to *p*-type and forming a lateral *p-n* junction. This direct, in-situ patterning process requires no additional masks, split gates, or multiple TMD heterostructures, offering a simple yet versatile route to functional junction devices.

The quality and performance of the resulting *p-n* junction were confirmed by sMIM imaging and electrical characterization. Figure 2b shows the phase channel of dMIM/dV mapping and corresponding line profile of a *p–n* junction patterned on an hBN-encapsulated $WSe_2$ device on a 90 nm $SiO_2$ substrate, revealing a ~180° phase shift across the junction boundary. This result confirms the dual types of carriers for the well-defined *p-n* junction (see Supplementary Note 1). Electrical characterization (Supplementary Fig. S6) further verifies the rectification behavior. Figure 2c displays the gate-dependent $I_D$–$V_{DS}$ characteristics of a $WSe_2$ rectifier patterned on 50 nm $Al_2O_3$ using 1 keV e-beam at $V_{SET1}$ = -10V (*n*-doping) and $V_{SET2}$ = 20V (*p*-doping). To evaluate junction quality, we fitted the data using a modified Shockley diode equation[51,52]:

$$I_{DS} = \frac{nV_T}{R_S} W\left[\frac{I_0 R_S}{nV_T}\exp\left(\frac{V_{DS} + I_0 R_S}{nV_T}\right)\right] - I_0$$

where $V_T = k_B T/e$ is the thermal voltage at temperature $T$, $k_B$ is the Boltzmann constant, $e$ is the electron charge, $I_0$ is the reverse saturation current, W denotes the Lambert W function, $n$ is the ideality factor, and $R_S$ represents the series resistance from the electrode/$WSe_2$ contacts and doping regions. The extracted ideality factor of $n = 1.3$ at $V_G = 0$ V (inset of Fig. 2c) reflects the high quality of the directly pattened junction. Importantly, the rectification ratio is highly gate-tunable, varying from $1.51 \times 10^4$ at $V_G = 0$ V to $9.18 \times 10^2$ at 3.5 V (Fig. 2d), illustrating the device's adaptability for dynamic circuit applications.

**Transistor performance optimization by e-beam-enabled band engineering**

Building on the capability of selective $p$- and $n$-doping via e-beam, we further demonstrate programmable band engineering in $WSe_2$ channels for transistor performance optimization. As illustrated in Fig. 3a, the entire channel was first uniformly doped into $n$-type (or $p$-type) via 1 keV e-beam exposure at $V_G = V_{SET1}$, followed by a second e-beam exposure in the central region under a different bias $V_{SET2}$ to form a local tunable energy barrier in the middle of the channel. sMIM-Im mapping and the corresponding line profile of a e-beam-engineered $WSe_2$ channel (Fig. 3b) confirm the modulation of conductivity, with a ~ 500 nm-wide region of suppressed conductance indicating the engineered potential barrier.

This spatially defined band modulation markedly improves the switching behavior of $WSe_2$ FETs. Figure 3c compares the transfer characteristics of a hBN-encapsulated $WSe_2$ device on 10-nm $HfO_2$ before and after band modulation. This pristine channel exhibits $n$-type behavior with a subthreshold swing (SS) of 169 mV/dec. Following modulation implemented via 1 keV e-beam with $V_{SET1}$ = -8 V and $V_{SET2}$ = 2 V, the same device exhibits a near-ideal SS value of ~ 60 mV/dec across four decades of current, representing a substantial improvement. Although e-beam exposure can reduce interface trap densities and slightly improve SS[50], this mechanism alone cannot fully account for the dramatic enhancement observed here. Instead, uniformly doped devices exhibit only a moderate SS reduction as shown in Supplementary Fig. S7. Therefore, the formation of a well-defined potential barrier via spatially controlled doping is the dominant mechanism for the observed SS optimization. Furthermore, this band engineering approach is robust and generally applicable to $WSe_2$ FETs on different dielectric substrates like $Al_2O_3$ with statistical repeatability (Supplementary Fig. S8).

To quantitatively understand observed SS improvement, we model the device as a Schottky barrier FET (SB-FET) within the well-established Landauer framework[53,54]. In the subthreshold regime where carrier scattering is negligible, the source-drain current per unit width $I_{DS}$ is expressed as:

$$I_{\mathrm{DS}} = \frac{2e}{h}\int_{\mathrm{E_c}}^{+\infty} T(E)M_{\mathrm{c}}(E)f(E)\mathrm{d}E$$

Here, $T(E)$ is the energy-dependent transmission probability across the Schottky barrier, $M_c(E)$

represents the density of conduction modes in the 2D channel, $f(E)$ is the Fermi-Dirac distribution, and $E_c$ defines the energy offset between the conduction band minimum and the quasi-Fermi level ($E_f$). Analytical expressions for these terms are detailed in Supplementary Note 2. The gate-voltage $V_G$ tunes $E_c$, thereby modulating both thermionic ($I_{thermal}$) and tunneling ($I_{tunneling}$) current components.

In the pristine channel modeled as a conventional SB-FET (Fig. 3d), the flat-band voltage $V_{FB}$ is defined as the gate bias at which $E_c$ aligns with the Schottky barrier height $\Phi_{SB}$ (Supplementary Fig. S9a). For $V_G < V_{FB}$, $I_{thermal}$ dominates, yielding an ideal SS of ~ 60 mV/dec, though typically beyond practical measurement limit. When $V_G$ surpasses $V_{FB}$, the increase of $I_{thermal}$ suddenly slows down as $\Phi_{SB}$ remains constant due to Fermi-level pinning. The total current is dominated by $I_{tunneling}$. However, the increase of $I_{tunneling}$ is limited by the fixed Schottky barrier width, ultimately degrading SS. Numerical simulations based on this model predict an SS of ~ 169 mV/dec (bottom panel of Fig. 3d and open circles in Fig. 3c), consistent with measured SS value in the pristine channel.

In contrast, the e-beam-modulated channel introduces a reconfigurable band profile (Fig. 3e). The initial global doping (set by $V_{SET1}$) narrows the Schottky barrier and enhances tunneling probability, accelerating the rise of $I_{tunneling}$. Simultaneously, the central energy barrier (defined by $V_{SET2}$) shifts the flat-band condition to $V_{FB}'$ and therefore confines tunneling to a narrower region when $V_G > V_{FB}'$, resulting in a steeper turn-on slope (Supplementary Fig. S9b). Applying the same Landauer formalism, our calculations predict a near-ideal SS of ~ 60 mV/dec (bottom panel of Fig. 3e and open circles in Fig. 3c), in excellent agreement with experiment.

This band-engineering approach is effective for both *p*- and *n*-type channels, enabling fine-tuning of threshold voltages and optimization of SS. To validate this, we fixed $V_{SET1}$ and varied $V_{SET2}$ during modulation of a $WSe_2$ device on a 10 nm $Al_2O_3$ dielectric. Figure 3f and 3g show the transfer curves of band-modulated *p*-FET and *n*-FET configurations, revealing systematic threshold voltage shifts and consistently optimized SS near 60 mV/dec. The extracted threshold voltages $V_{TH}$ exhibit linear dependence on $V_{SET2}$ (Fig. 3h), matching theoretical predications from our band-engineering model (open circles in Fig. 3f, g; see also Supplementary Note 2). This electrical tunability of $V_{TH}$, combined with ideally optimized SS in both polarities, offers a robust platform for designing and constructing on-demand high-performance, rewritable CMOS logic circuits.

**Reconfigurable high-performance CMOS logic devices via direct e-beam writing**

To demonstrate reconfigurable complementary logic functionalities through direct e-beam writing, we fabricated several rewritable $WSe_2$ units on 50 nm $Al_2O_3$ substrates (Supplementary Fig. S10a), where the thicker dielectric enhances *p*-type doping efficiency. In contrast to previous $WSe_2$ CMOS implementations that require specialized contact treatments[9,20,29–31,42–44], our e-beam band-engineering approach enables the realization of high-performance *p*- and *n*-FETs in adjacent channels using standard Cr/Au electrodes, forming CMOS inverters (Fig. 4a and 4b). Supplementary Fig. S10b presents the transfer curves of e-beam patterned *p*- and *n*-FETs, showing slight SS degradation likely due to increased interface traps in the thicker oxide.

By tuning the e-beam writing parameters, we can continuously optimize key inverter characteristics by multi-$V_{TH}$ optimization, including operating voltage, voltage gain, and power consumption. As illustrated in Fig. 4c, simultaneous leftward shifting of the threshold voltages of both *p*- and *n*-FETs via band modulation allows the reduction of the inverter's operating voltage. We implemented this strategy by successively rewriting a CMOS inverter using $V_{SET2}$ values from 1.5 V to 0.25 V. Figure 4d shows the resulting voltage transfer curves measured at a fixed $V_{DD}$ = 0.5 V, demonstrating a clear reduction in switching voltage with decreasing $V_{SET2}$. Furthermore, by fixing one channel and modulating the other, the intersection point of the *p*- and *n*-channel transfer curves can be finely adjusted (Fig. 4e), enabling precise control over voltage gain and power consumption. Detailed demonstration is provided in Supplementary Fig. S11. Figure 4f shows the optimized CMOS inverters written on the same device, achieving peak voltage gains of 25.7, 50.8, and 73.1 at supply voltages of $V_{DD}$ = 1.0 V, 1.5 V and 2.0 V, respectively. The corresponding peak power consumptions are 0.14 nW, 2.0 nW and 3.4 nW (Fig. 4g), exceptionally low due to the advantages of CMOS design in combination with our e-beam-enabled fine-tuned optimization. This approach offers a simple, efficient and flexible platform for on-demand functional enhancement of CMOS logic devices without physical reconstruction or additional fabrication steps, which are typically impractical in conventional methods.

Beyond in-situ functional optimization, our technique also allows full reconfigurability of logic functions using the same rewritable units. To demonstrate this, we first reversed the polarity of $WSe_2$ channels via e-beam rewriting, as illustrated in Fig. 5a and 5b. Initially two adjacent channels were patterned into optimized *n*- and *p*-type configurations, forming an inverter with the corresponding transfer characteristics shown in Fig. 5a and its inset. We then rewrote these channels to reverse their polarities, producing an inverter with similar functionality but opposite doping configurations (Fig. 5b and inset). Leveraging this reconfigurability, we further programmed a single device comprising four adjacent rewritable channels to toggle between NAND and NOR logic functions without altering the layout or interconnects. A NAND gate was first realized by patterning the four channels into *p*, *n*, *n*, *p* configurations (Fig. 5c), equivalent to integrating two e-beam-defined inverters. The NAND functionality was confirmed by the measured input/output waveforms under $V_{DD}$ = 2 V, as shown in Fig. 5d. Subsequently, we inverted the polarity of each FET channel via e-beam rewriting and swapped the $V_{DD}$/GND terminals, yielding a NOR gate on the same physical structure (Fig. 5e), as verified by the measured output shown in Fig. 5f. This seamless logic transition, achieved without any physical modification, demonstrates the versatility and utility of rewritable 2D CMOS units for adaptive prototyping and rapid exploration of logic circuits and functions.

**Conclusion**

In summary, we have established a rewritable 2D electronics platform based on e-beam programmable ambipolar $WSe_2$ FETs, enabling site-selective, reversible, and high-resolution control over both *n*- and *p*-type doping. This approach allows wide-range carrier control for both polarities and precise band engineering within individual channels, yielding in-situ transistor optimization with near-ideal SS (~ 60 mV/dec) and finely tunable threshold voltages for both polarities. These capabilities enable on-demand balanced ambipolar performance and multi-$V_{TH}$ optimization for high-performance CMOS logic units, including inverters with high voltage gains and low power consumption. Moreover, the ability to seamlessly reconfigure logic functions, such as NAND-to-NOR transitions, through direct rewriting on the same physical layout provides unprecedented flexibility for rapid prototyping and functional adaptation in 2D complementary logic architectures.

Unlike traditional doping or contact engineering methods, our e-beam-enabled strategy is inherently scalable, mask-free, and fully compatible with CMOS back-end processes. Looking ahead, this method opens new pathways for developing rewritable neuromorphic circuits, adaptive electronics, and secure hardware platforms. By integrating with wafer-scale 2D material synthesis and parallel e-beam systems, our approach could evolve into a versatile and scalable framework for next-generation reconfigurable electronics.

## Methods

### Substrate preparation and device fabrication

285 nm $SiO_2$/Si substrates were purchased from BONDA TECHNOLOGY. Other substrates with 90 nm $SiO_2$, 50 or 10 nm $Al_2O_3$, 50 or 10 nm $HfO_2$ were prepared by thermally growing $SiO_2$ on Si wafers or depositing $Al_2O_3$/$HfO_2$ via atomic layered deposition (ALD, Picosun R200) onto 285 nm $SiO_2$/Si wafers. For selected devices, pre-patterned Au bottom gates were fabricated on $SiO_2$ substrates via thermal evaporation prior to ALD dielectric deposition.

High-quality BN crystals (from K. Watanabe and T. Taniguchi) were exfoliated onto 285 nm SiO2/Si substrates. BN flakes with thickness of 10 - 20 nm were selected for both top and bottom encapsulation, enabling balanced e-beam induced *n*-doping and *p*-doping behavior. BN-encapsulated $WSe_2$ heterostructures were assembled on various substrates using standard dry transfer technique[55], followed by annealing in Ar/$H_2$ forming gas at 350 °C for 3 hours.

Standard electron-beam lithography was used to define etching masks and electrode patterns. All devices employed thermally evaporated Cr/Au contacts (3 nm/70 nm). For reconfigurable logic gate demonstrations, reactive-ion etching (RIE) was used to define four isolated channels on a single $WSe_2$ device with multiple electrodes.

### E-beam doping and electrical measurements

Devices were mounted in a SEM (TESCAN VEGA LMS) using a custom stage with electrical feedthroughs for in-situ e-beam doping and electrical measurements. Transport properties were measured using a Keithley 2636A source meter in the SEM chamber under a vacuum of 0.075 Pa. Unless otherwise noted, e-beam doping was performed with 1 keV energy and 10 pA beam current. Both normal scanning and lithography modes were employed to write and investigate pre-designed doping patterns on BN-encapsulated $WSe_2$ devices.

For *p*-*n* junctions created on 50 nm $Al_2O_3$ substrate, the full channel was first exposed to the e-beam (1 keV, 10 pA) for 60 s in normal scanning mode under gate bias $V_G$ = -10 V to induce *n*-doping. Subsequently, half of the channel was selectively scanned under $V_G$ = 25 V for 30 s to create a *p*-doping region.

Band-modulated *p*-FETs and *n*-FETs were programmed by first globally doping the channel under a positive $V_G$ (8 V for 10 nm $Al_2O_3$, 25 V for 50 nm $Al_2O_3$) for *p*-FETs or a negative $V_G$ (-8 V for 10 nm $Al_2O_3$, -10 V for 50 nm $Al_2O_3$, -8 V for 10 nm $HfO_2$) for *n*-FETs. Then the scanning window was confined to the center of the channel with $V_G$ swept between -8 V and +8 V to fine-tune the barrier profile.

For logic gates, channels were interconnected using BNC cables according to the desired logic gate configuration, and the operating $V_{DD}$ was applied during doping. Each channel was first globally doped to set its polarity, followed by simultaneous band modulation process across all channels under the same $V_G$ to achieve high-device performance.

**sMIM characterization**

sMIM was performed on a commercial Bruker AFM system (Dimension Icon) at room temperature. The technique utilizes a cantilever-based AFM combined with a 3 GHz microwave signal delivered through a customized shielded cantilever probes commercially available from PrimeNano Inc. Two measurement schemes were employed in this study, i.e., direct sMIM imaging and differential sMIM imaging. In direct sMIM imaging, a 3 GHz microwave signal is locally shined on the sample surface and the reflected microwave signal is collected. This reflected signal is further demodulated into two orthogonal channels, the sMIM-Im and sMIM-Re. sMIM-Im increases monotonically as the sample's conductivity increases, and is commonly used to characterize the sample's local conductivity as shown in Fig. 1(a) and 3(b). In differential sMIM imaging, an oscillating voltage bias can be applied to the tip to generate an AC electric field to modulate the samples' electrical properties. The resulting differential sMIM signals would represent dMIM/dV. For semiconductors, such dMIM/dV measurement is highly useful as the amplitude (phase) of the dMIM/dV signal characterizes the carrier doping level (type) of the semiconductor. In this study, we use the phase channel of dMIM/dV mapping (Fig. 2b) to characterize the spatial doping profile of the *p-n* junction. Some doping patterns for sMIM imaging were created by e-beam exposure at 2 keV with 10 pA beam current.

**Theoretical simulation**

The e-beam-defined doping generates local electrostatic potentials and enables precise band modulation of the transistor channel, which can be quantitatively simulated. To model the transport behavior of the band-modulated channel, we employed the Landauer formalism[53,54]. The drain current at a given gate voltage was calculated by numerically integrating the Landauer equation, incorporating both thermionic and tunneling contributions. All simulations were carried out using Wolfram Mathematica 11.3, with further details provided in Supplementary Note 2.

**Acknowledgements**

W. S. acknowledges support from National Key Research and Development Program of China (Grant No. 2024YFA1409003 and No. 2024YFB3614103) and National Natural Science Foundation of China (Grant No. 12274090 and Grant No. 12574187). X.D.Z. acknowledges support from National Key Research Program of China (Grant No. 2024YFA1409003) and National Natural Science Foundation of China (Grants No. 12350401 and 12274088). J. S. acknowledges support from National Key Research Program of China (Grant No. 2022YFA1403300), the Innovation Program for Quantum Science and Technology (Grant No. 2024ZD0300103) and Shanghai Municipal Science and Technology Major Project (Grant No. 2019SHZDZX01). Part of the sample fabrication was performed at the Fudan Nanofabrication Laboratory.

**Author contributions**

W.S. and A.Z. conceived the project. T.T. and K.W. provided the hBN crystals. J.W. and J.L.W. prepared the $Al_2O_3$ and $HfO_2$ substrates. Q. L., W.Q.S., L.W., J.Q.L., and Z.J.G. fabricated the devices and performed electrical measurements. S.Y.Z., Y.Z., and X.D.Z. conducted the TEM characterization. W.S., Q.L., W.Q.S, H.H., J.L.W., J.S. and X.D.Z. analyzed the data, Q.L. carried out theoretical calculations. W.S., Q. L., W.Q.S., and X.D.Z. wrote the manuscript with input from all authors.

**Competing interests**

The authors declare no competing interests.

## Figures

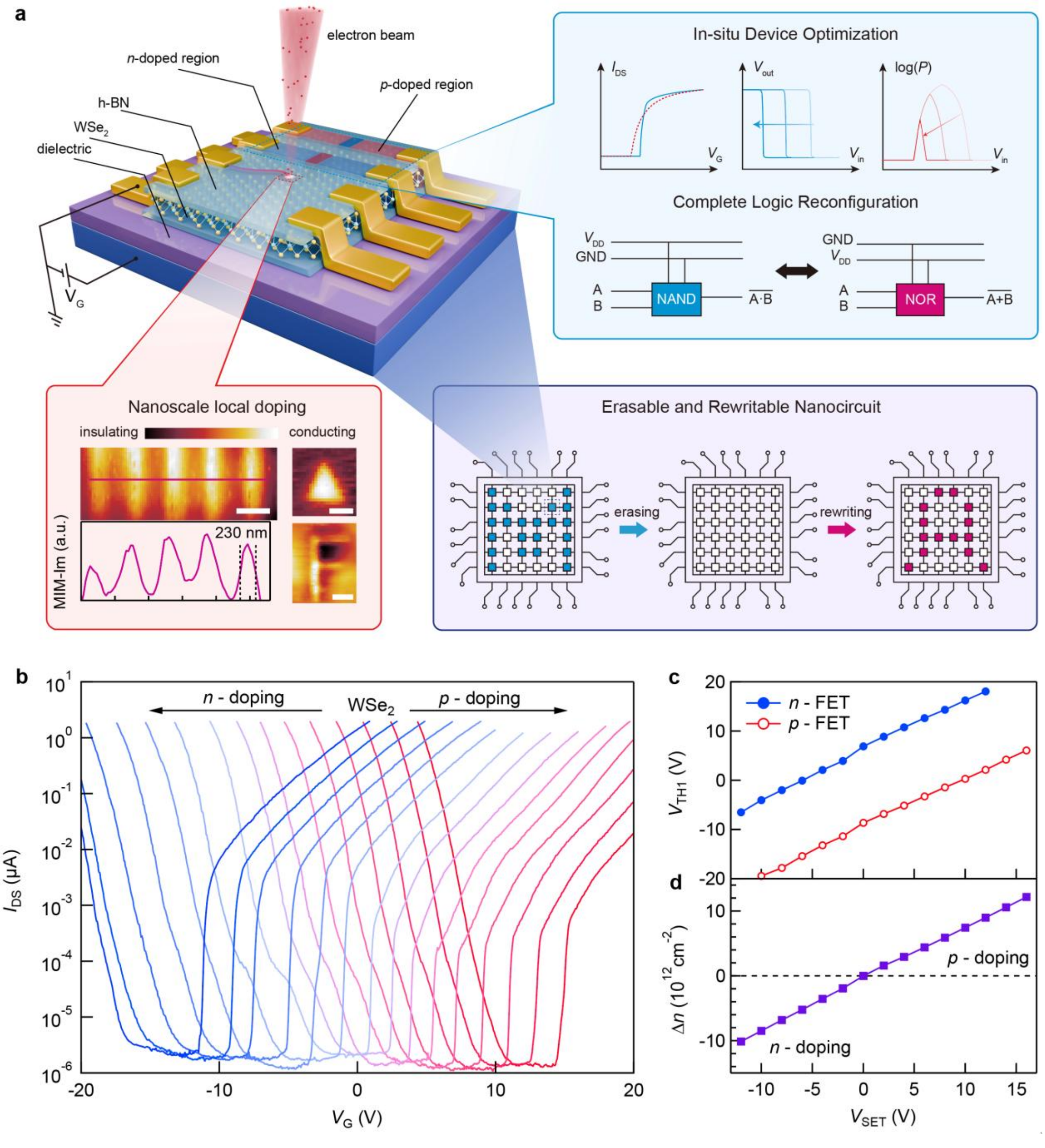

**Fig. 1 | Rewritable electronics platform based on ambipolar $WSe_2$ FETs enabled by controllable e-beam doping. a,** Schematic of the e-beam-programmable FET platform, enabling high-resolution, nonvolatile doping with reversible polarity control. The nanoscale written doping patterns are visualized by sMIM, revealing high spatial resolution of ~ 200 nm line widths. Scale bar: 500 nm ((line array), 500 nm (triangle) and 1 μm (letter). This capability supports in-situ device optimization and complete logic reconfiguration, offering a scalable

route for rewritable nanocircuits and rapid prototyping. **b,** Transfer curves ($I_{DS}$-$V_G$) of an hBN-encapsulated $WSe_2$ FET on 50 nm $Al_2O_3$, measured at $V_{DS}$ = 0.1 V after sequential 1 keV e-beam exposures across the entire channel at different set voltages $V_{SET}$ from -10 V to 16 V (left to right, step: 2 V). Arrows indicate nonvolatile *p*- and *n*-doping reversibly controlled by $V_{SET}$. **c,** Threshold voltage $V_{TH}$ as a function of $V_{SET}$ for *n*-type and *p*-type FET operation in the same device, both showing linear dependence on $V_{SET}$. **d,** Corresponding carrier density change $\Delta n$ for *p*- and *n*-doping in the same devices, confirming precise controllability of the e-beam doping approach.

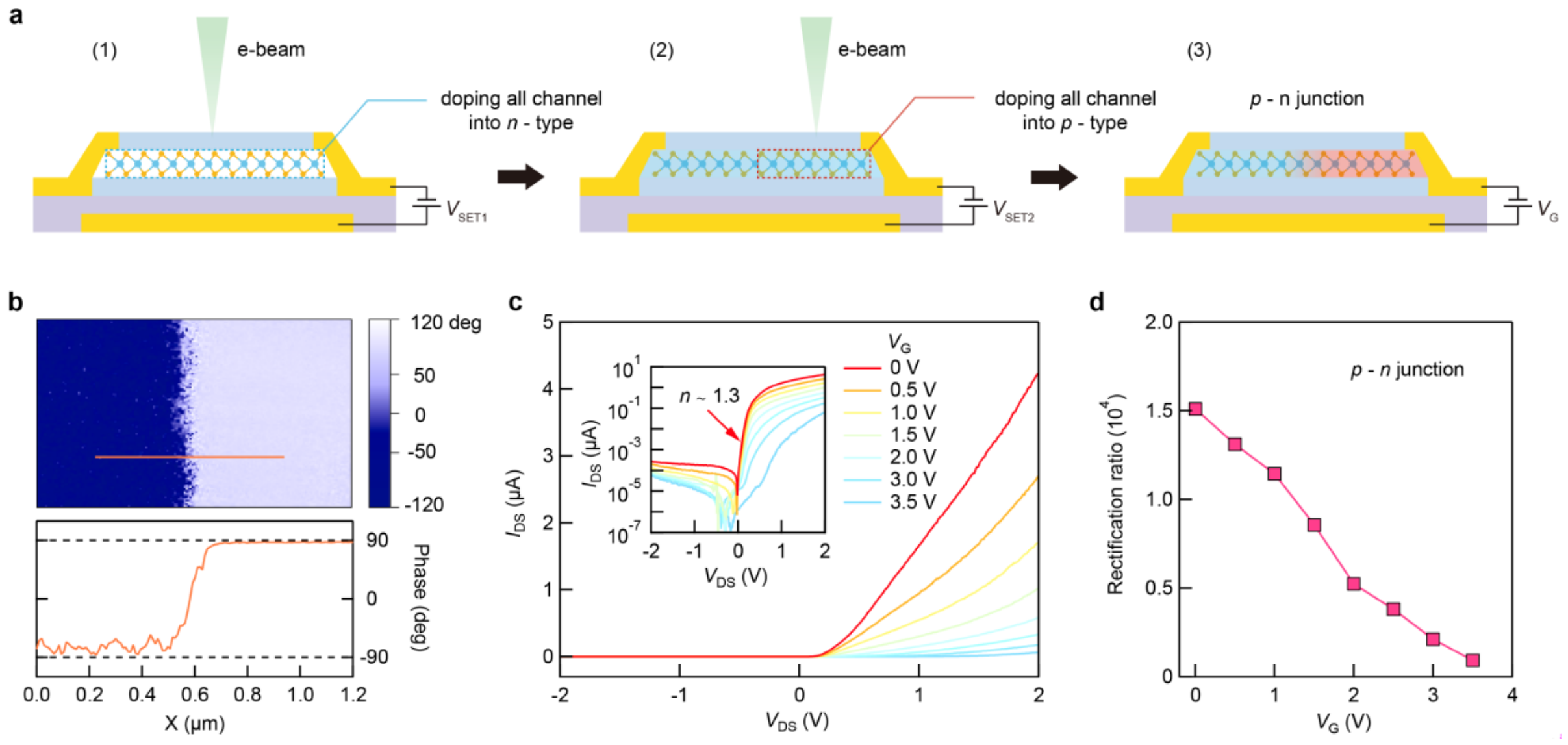

**Fig. 2 | Gate-tunable rectification in a rewritable $WSe_2$ FET with e-beam-defined *p-n* junction. a,** Schematic of spatially selective e-beam doping to pattern a lateral *p–n* junction within a single $WSe_2$ device. The entire channel is first globally doped *n*-type using 1 keV e-beam at $V_G$ = $V_{SET1}$, followed by selective re-doping of half the channel to *p*-type under $V_{SET2}$, forming a reconfigurable lateral *p-n* junction. **b**, The phase channel of dMIM/dV mapping (top) and corresponding line profile (bottom) showing a sharp 180 degree phase shift across the junction, indicative of a well-defined *p-n* boundary. **c**, Gate-tunable current-voltage ($I_{DS}$-$V_{DS}$) characteristics of the e-beam defined $WSe_2$ *p–n* junction on a 50 nm $Al_2O_3$ substrate. Inset: Log-scaled plot of $|I_{DS}|$ versus $V_{DS}$, showing an ideality factor of ~ 1.3 at $V_G$ = 0 V. **d**, Gate-dependent rectification ratio (forward/reverse current) derived from the data in **c**.

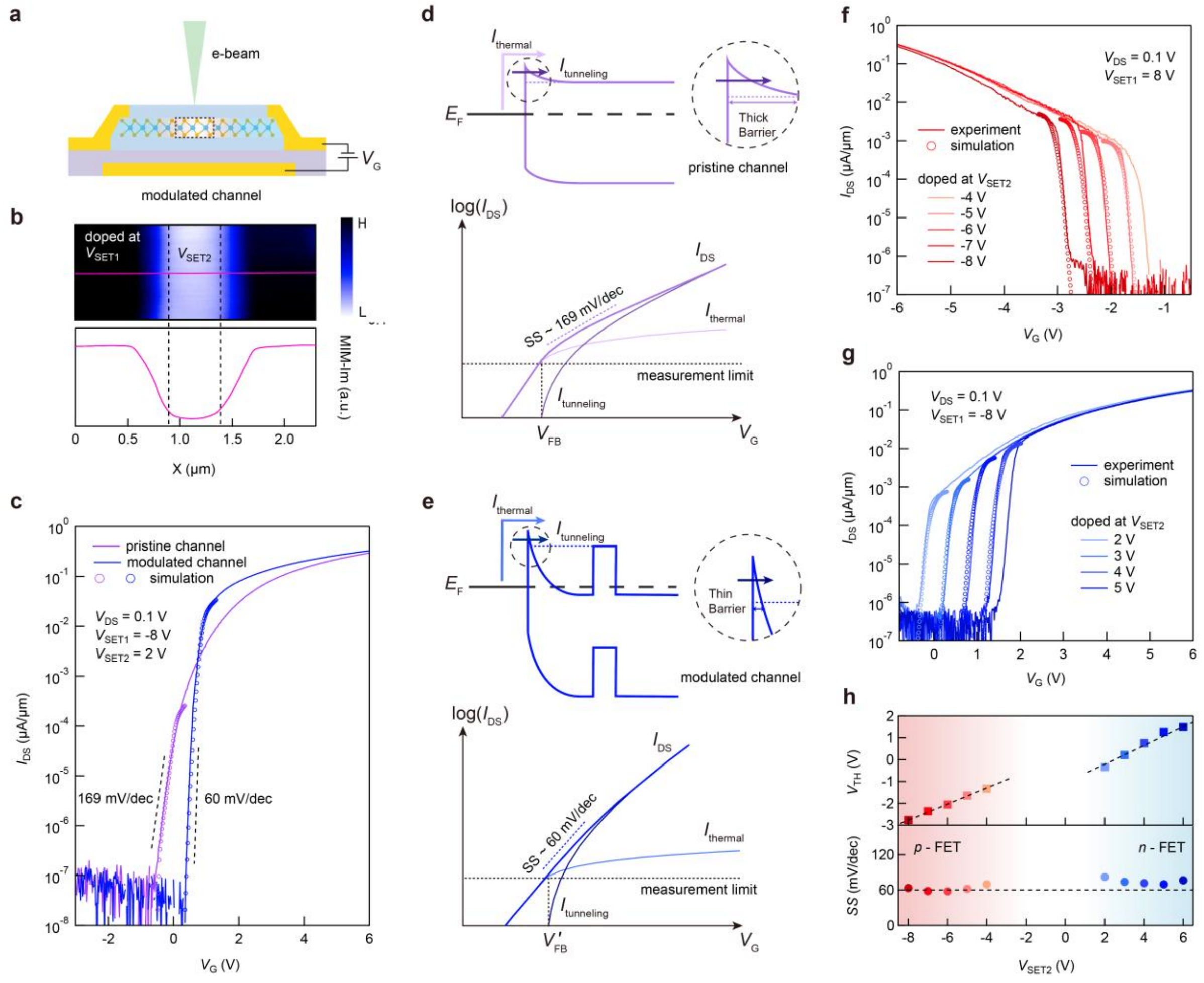

**Fig. 3 | FET performance optimization via e-beam-enabled band modulation in rewritable $WSe_2$ devices. a,** Schematic illustration of spatially resolved band modulation in rewritable $WSe_2$ FET by site-selective e-beam doping. **b,** sMIM imaging and corresponding line profile of a band-modulated *n*-channel in $WSe_2$ on a 90 nm $SiO_2$ substrate. The full channel was initially doped into the desired polarity using 1 keV e-beam at $V_G = V_{SET1}$, followed by re-doping the central region at $V_{SET2}$ to introduce a tunable energy barrier for performance optimization. **c**, Transfer characteristics ($I_{DS}$ - $V_G$) of a rewritable $WSe_2$ FET on 10 nm $HfO_2$ before (pristine) and after band modulation, showing improved switching behavior with near-ideal SS of ~ 60 mV/dec over four orders of magnitude in $I_{DS}$. **d, e,** Energy band diagrams (top) and simulated current components (bottom) of the pristine channel (**d**) and band-modulated channel (**e**), illustrating how the engineered barrier enhances tunneling currents to improve subthreshold characteristics. **f, g,** Transfer characteristics of band-modulated *p*-type (**f**) and *n*-type (**g**) FETs in a single $WSe_2$ device on 10 nm $Al_2O_3$ with fixed $V_{SET1}$ and variable $V_{SET2}$, demonstrating systematic threshold voltage shifts and optimized transistor behavior. Open

circles denote simulated curves based on the band-modulation transport mechanism. **h,** Extracted threshold voltages and SS values of the modulated *p*-FET and *n*-FET, highlighting precise control over threshold voltages via $V_{SET2}$ while maintaining optimal SS values near 60 mV/dec.

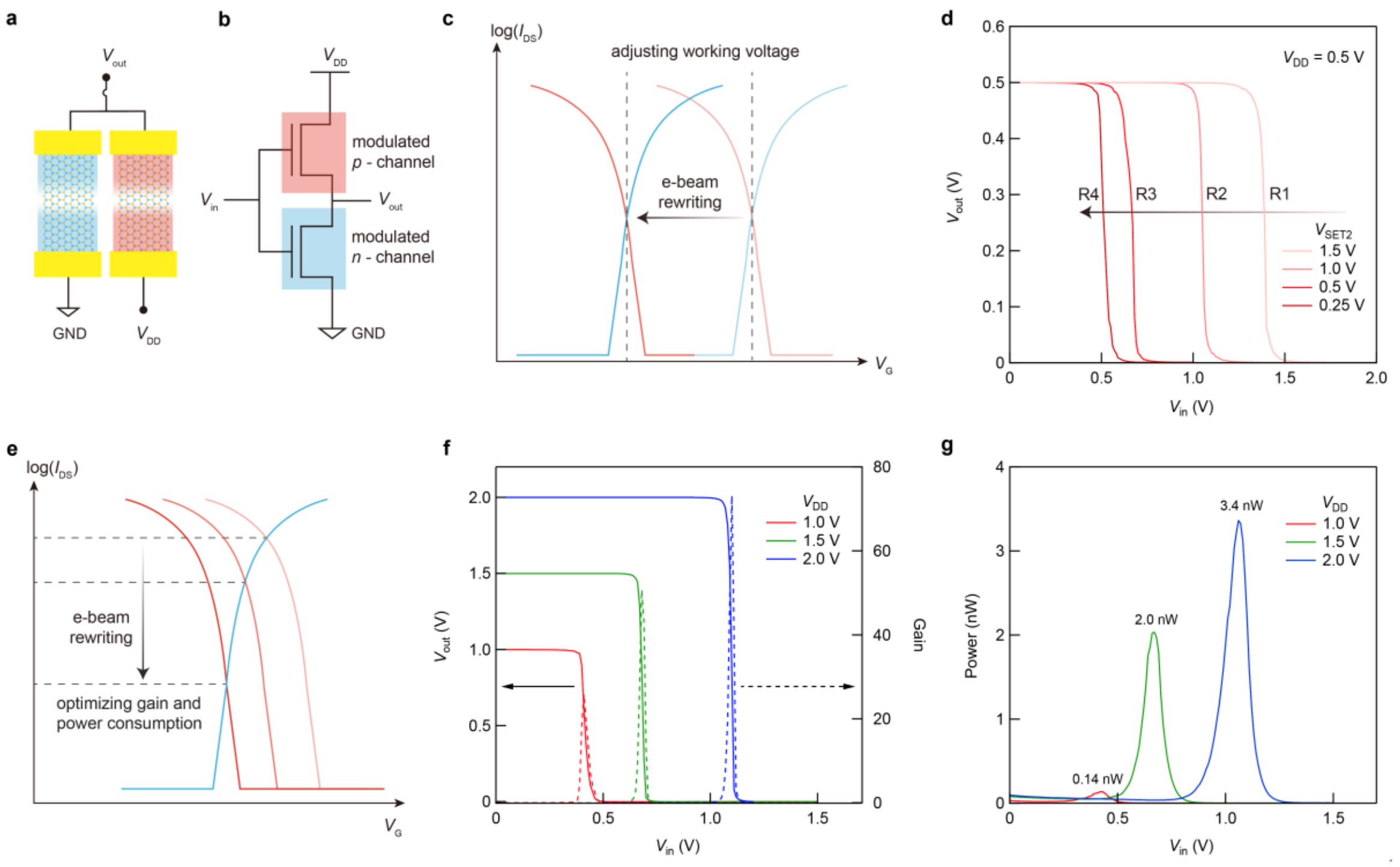

**Fig. 4 | High-performance CMOS inverter enabled by band-modulated rewritable $WSe_2$ FETs. a,** Schematic of a rewritable CMOS inverter composed of two $WSe_2$ channels in a single device configured as modulated *p*-FET and *n*-FET through site-selective e-beam doping. **b,** Circuit diagram of the CMOS inverter based on the rewritable *p*- and *n*-type FETs. **c,** Conceptual illustration of adjusting the inverter's operating voltage by e-beam band modulation to shift the threshold voltages simultaneously. **d,** Voltage transfer characteristics of the rewritable CMOS inverter after sequential rewritten from R1 to R4 with different $V_{SET2}$, demonstrating tunable switching voltage via controlled band modulation. **e,** Illustration of the optimization principle for the CMOS inverter. By fixing one channel while modulating the other, the intersection point of the *p*- and n-channel transfer curves shifts, allowing fine-tuning of voltage gain and power consumption. **f,** Voltage transfer characteristics of optimized CMOS inverters operated at various supply voltages ($V_{DD}$ = 1.0, 1.5, 2.0 V), yielding peak voltage gains of 25.8, 50.7, and 73.1, respectively. **g,** Corresponding power consumption of the

inverters at each $V_{DD}$, confirming low-power operation.

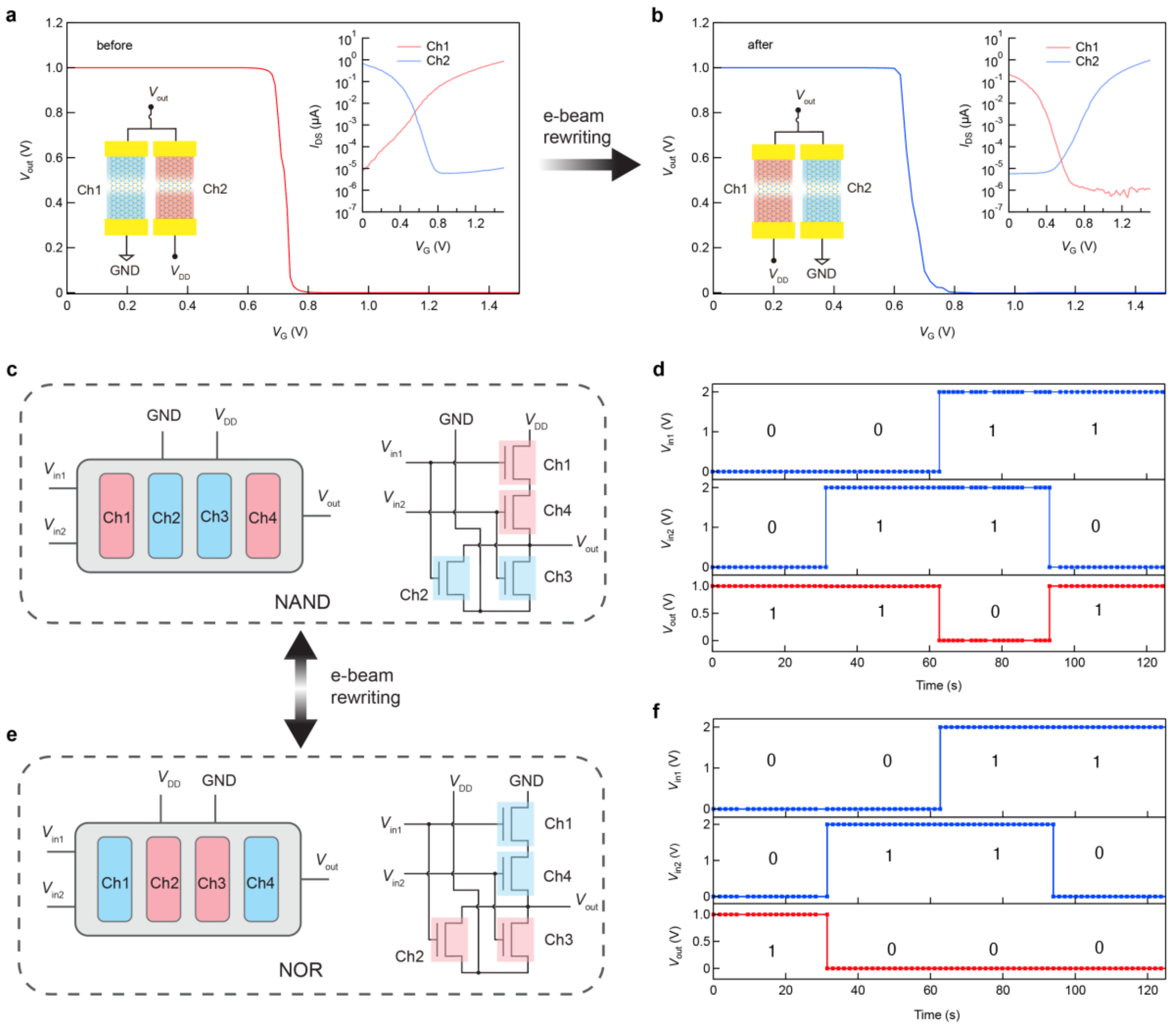

**Fig. 5 | Logic gate reconfiguration from NAND to NOR enabled by rewritable $WSe_2$ FETs.** **a, b,** Inverter characteristic of two neighboring $WSe_2$ channels before (**a**) and after (**b**) polarity reprograming through e-beam rewriting in a single device. All channels are patterned or reconfigured into *p*- or *n*-type via selective e-beam band modulation. Insets show the schematic and corresponding transfer curves of the individual channels, confirming successful and reversible polarity control. **c,** Schematic of a NAND logic gate implemented using four neighboring rewritable $WSe_2$ units on a single device (left), and its corresponding circuit diagram (right). **d,** Measured input/output voltage waveforms of the NAND gate under $V_{DD}$ = 2V, demonstrating correct logic operation. **e,** Schematic of the NOR logic gate reconfigured from the same four $WSe_2$ units by reversing their polarities via e-beam rewriting (left), along with the underlying circuit diagram (right). **f,** Measured input/output voltage waveforms of the

reconfigured NOR gate at $V_{DD}$ = 2V, confirming successful on-device logic transition without physical reconstruction.